\documentclass[aps,prb,twocolumn]{revtex4}  
\usepackage{graphicx}
\usepackage{dcolumn}
\usepackage{bm}
\usepackage{amsmath}
\newcommand{\comment}[1]{}

\begin{document}
\renewcommand{\theequation}{\arabic{section}.\arabic{equation}}


\title{Direct Measurement of the Surface Tension of Nanobubbles}

\author{Phil Attard}

\affiliation{\emph{phil.attard1@gmail.com}, Sydney, Australia}

\date{9 May, 2015}

\begin{abstract}
It is shown that when the nanobubble contact line is pinned to a penetrating tip
the interface behaves like a Hookean spring
with spring constant proportional to the nanobubble surface tension.
Atomic force microscope (AFM) data for several nanobubbles and solutions
are analysed and yield surface tensions in the range 0.04--0.05$\,$N/m
(compared to 0.072$\,$N/m for saturated water),
and supersaturation ratios in the range 2--5.
These are the first direct measurements of the surface tension
of a supersaturated air-water interface.
The results are consistent with recent theories
of nanobubble size and stability,
and with computer simulations of the surface tension
of a supersaturated solution.
\end{abstract}

\maketitle

%
\section{Introduction}
\setcounter{equation}{0} 
%


Experimental evidence for nanobubbles was first published in 1994,
\cite{Parker94}
and since then their existence has been confirmed
from various features of the measured forces
between hydrophobic surfaces,
\cite{Carambassis98,Tyrrell01,Tyrrell02}
including a reduced attraction in de-aerated water,
\cite{Wood95,Meagher94,Considine99,Mahnke99,Ishida00,Meyer05,Stevens05}
and from images obtained
with tapping mode atomic force microscopy.
\cite{Tyrrell01,Tyrrell02,Ishida00,Holmberg03,Simonsen04,%
Zhang04,Zhang06a,Zhang06b,Yang07}
For a recent review of theory and experiment,
see Ref.~\onlinecite{Attard14}.

The initial controversies over the existence of nanobubbles
---that they should have an internal gas pressure of 10--100 atmospheres
that would cause them to dissolve in microseconds---
have largely been resolved.
First it was shown that nanobubbles can only be in equilibrium
in water supersaturated with air,\cite{Moody02}
and then it was shown that the surface tension of a supersaturated solution
must be less than that of a saturated solution.\cite{Moody03}
Finally, it was shown that nanobubbles with a pinned contact rim
are thermodynamically stable.\cite{Attard15}
This means that contact line pinning
is a necessary and sufficient condition
for nanobubbles to be in mechanical and diffusive equilibrium with the
supersaturated solution.

The significance of the reduction in surface tension
is that the internal gas pressure,
which can be calculated from the Laplace-Young equation,
is much less than those initial estimates
used to argue against nanobubbles.
It also means that the  degree of supersaturation of the solution,
necessary for diffusive equilibrium of the nanobubble,
is reduced to realistic levels that are attainable in the fluid cell.

For many it is surprising that purely on thermodynamic grounds
(ie.\ no additives or surfactant)
the surface tension for nanobubbles should be reduced from
the usual value of the air-water interface.
Nevertheless, this result is firmly established by
thermodynamics,\cite{Moody03}
density functional theory,\cite{Cahn59,Oxtoby88,Oxtoby98}
and computer simulation.\cite{Thompson84,Moody04,He05}
The result is not widely acknowledged within the nanobubble field,
possibly because many practitioners place more trust in experimental measurement
than they do in  thermodynamics or in mathematical equations.
To close this gap,
it would be desirable to measure directly the surface tension
of nanobubbles.

Measuring the surface tension of the supersaturated
air-water interface is a worthwhile experimental goal
that has application beyond nanobubbles.
For example,
in atmospheric physics the nucleation of cloud droplets
always occurs from a  supersaturated atmosphere,
and so the rate of change of surface tension
with the degree of supersaturation
is a key input necessary for quantitative modeling in that field.

\begin{figure}[t!]
\centerline{
\resizebox{8.5cm}{!}{ \includegraphics*{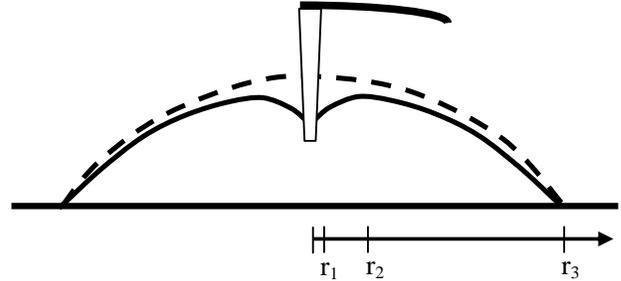} } }
\caption{\label{Fig:geom}
Hemispherical bubble penetrated and deformed by a conical tip
(blunt radius $r_0$, height $z_0$)
upon which it sticks at $r_1$.
}
\end{figure}

A supersaturated solution in the current context means
that the concentration of air in the water is higher
than it would be if the water were equilibrated with air
at the current temperature and pressure.
Supersaturation can be readily achieved by previously
equilibrating the water with air at a higher pressure
or at a lower temperature.
The difficulty in measuring directly
the surface tension of a supersaturated solution
is that the change in surface tension is determined by the concentration
of air within the first nanometer of the interface,
but, due to diffusion across the interface,
this region is always saturated rather than supersaturated,
at least if the measurement is performed on a macroscopic droplet or bubble
exposed to the atmosphere.
In the case of nanobubbles, however,
the solution is supersaturated immediately in the vicinity of the interface
due to the high internal gas pressure of the nanobubble,
and so the surface tension of a nanobubble must be that
of a supersaturated solution.
If one could measure the surface tension of the nanobubble,
then one would have a means to determine
the rate of change of surface tension with supersaturation.
The present author knows of no previous measurements
of the surface tension of the supersaturated air-water interface.

The present paper is concerned with modeling atomic force microscope (AFM)
force measurements on nanobubbles when the tip of the cantilever
penetrates the nanobubble.
The aim is to relate the surface tension of the nanobubble
to the measured force.
The results are applied to force data from several nanobubbles.
In all cases analyzed, the surface tension was always less
than that of the saturated air-water interface,
and the supersaturation ratio of the solution
was always greater than unity.

\section{Penetrated Hemispherical Bubble}
\setcounter{equation}{0} \setcounter{subsubsection}{0}
%

Figure~\ref{Fig:geom} is a sketch of a hemispherical bubble on a solid
surface that is deformed by a penetrating conical tip.
In a cylindrical coordinate system,
the bubble  contacts
the surface at $(r_3,0)$ and the tip at $(r_1,z_1)$.
The bubble is pinned or fixed at these two radii.
The blunt tip of the cone is at a height (separation) above
the substrate of $z_0$ for $r \le r_0$.
The half angle is given by $\tan \alpha = (r_1-r_0)/(z_1 - z_0 )$.
The case of a positive load, $F>0$, is shown in the figure;
for a negative force the bubble is extended and the dimple vanishes.

The  following analysis is similar to that given earlier
in that it is based on a small force expansion.\cite{Attard01}
The earlier analysis explicitly included
the effects of a surface or interaction force between the probe and the bubble,
whereas here the bubble is penetrated by the probe and makes
either stick or slip contact with its sides.
The earlier analysis was for a bubble mobile on the substrate
and for fixed number of air molecules,
whereas here the bubble is pinned at $r_3$
and is in diffusive equilibrium with the supersaturated solution.
Due to the present pinned contact rim,
the solid surface energies and the contact angle
play no role in the present analysis,
whereas they did for the case of a mobile bubble
analyzed in  Ref.~\onlinecite{Attard01}.
These differences in the model make the present analysis
considerably simpler and shorter than that of the earlier case.
Despite these differences in the model,
the qualitative conclusion is the same in both cases,
namely that the bubble interface behaves as a Hookean spring.
The quantitative expression for the spring constant
found here is of course specific for the present model
of a bubble pinned at $r_3$.

\subsection{Undeformed Bubble}

As recently shown,\cite{Attard15}
once the bubble is pinned at the contact radius $r_3$,
it is thermodynamically stable at the critical radius and density.
Accordingly, the undeformed bubble is the critical bubble.
It is of course hemispherical,
and its radius is the critical radius\cite{Attard15,Rowlinson82}
\begin{equation} \label{Eq:Rcrit}
R_\mathrm{c} = \frac{2 \gamma}{(s-1)p}  = \frac{2 \gamma}{\Delta_p} .
\end{equation}
Here $s>1$ is the supersaturation ratio,
(the solution is necessarily supersaturated with air
in order for the bubble to be in diffusive equilibrium),\cite{Moody02}
$\gamma$ is the liquid-vapor surface tension
(of the supersaturated interface),\cite{Moody03,Moody04,He05}
$p$ is the pressure of the reservoir
(taken to be atmospheric),
$\Delta_p$ is the excess pressure of the bubble,
$k_\mathrm{B}$ is Boltzmann's constant, and $T$ is the temperature.
For simplicity,
the reservoir pressure and the saturation vapor pressure
are approximated as equal.\cite{Attard15}

The  undeformed  bubble has volume\cite{Attard15}
\begin{equation} \label{Eq:Vc}
V_\mathrm{c} =
\pi \left[ R_\mathrm{c} z_\mathrm{c}^2 - z_\mathrm{c}^3/3 \right] ,
\end{equation}
liquid-vapor surface area
\begin{equation}
A_\mathrm{c} =
2 \pi R_\mathrm{c}
\left[ R_\mathrm{c} - \sqrt{R_\mathrm{c}^2 - r_3^2}   \right]
= 2 \pi R_\mathrm{c} z_\mathrm{c},
\end{equation}
and apex height
\begin{equation}
z_\mathrm{c}
= R_\mathrm{c} - \sqrt{R_\mathrm{c}^2 - r_3^2} .
\end{equation}
The undeformed profile is
\begin{equation} \label{Eq:zcr}
z_\mathrm{c}(r) =
z_\mathrm{c}
- R_\mathrm{c} + \sqrt{R_\mathrm{c}^2 - r^2} .
\end{equation}

The undeformed bubble contains
$ N_\mathrm{c} = s pV_\mathrm{c} /k_\mathrm{B}T$
gas molecules.
In the present case of fixed contact rim $r_3$,
thermodynamic stability holds simultaneously for number
and volume fluctuations.\cite{Attard15}
Hence one does not have to insist upon constant number
as the author has had to do in previous work.\cite{Parker94,Attard96,Attard00}

\subsection{Deformed Bubble: Prick Stick}

The bubble profile is $z(r)$,
which ends at the contact points $(r_1,z_1)$ and $(r_3,0)$.
The contact points on the blunt tip satisfy
$r_1-r_0 = [z_1-z_0] \tan \alpha$,
where $\alpha$ is the cone half angle,
and $z_0$ is the height of the tip of the tip above the substrate,
which is also called the separation.
The tip is taken to be perfectly blunt,
which is to say that its end is a disc of radius $r_0$ at $z_0$.
Although in principle one could also have contact at $(r_1,z_0)$
for $r_1 \le r_0$, this case will be excluded in the numerical results below.
A perfectly sharp tip has $r_0 = 0$.

The volume is
\begin{eqnarray}
V[z] & = &
2 \pi \int_{r_1}^{r_3} \mathrm{d}r \, r z(r)
+ \pi r_1^2 z_1 - V_\mathrm{t}
\nonumber \\ & = &
2 \pi \int_{r_1}^{r_3} \mathrm{d}r \, r z(r)
+ \frac{2 \pi r_1^3}{3 \tan \alpha}
\nonumber \\ && \mbox{ }
+ \pi r_1^2 \left[ z_0 -\frac{r_0}{\tan \alpha} \right]
- \frac{\pi r_0^3}{3 \tan \alpha} .
\end{eqnarray}
The volume of the blunt tip inside the bubble is
\begin{equation}
V_\mathrm{t} =
\frac{\pi r_1^3}{3\tan \alpha}
- \frac{\pi r_0^3}{3 \tan \alpha} .
\end{equation}
The terms $\pi r_1^2 z_1 - V_\mathrm{t}$
give the vapor volume beneath the tip, $r < r_1$.
These may be neglected for the profile differentiation
since they are constant.
The fluid-vapor  interfacial area is
\begin{equation}
A[z]
=
 2 \pi \int_{r_1}^{r_3} \mathrm{d}r \,r  \sqrt{1 + z'(r)^2}.
\end{equation}

The total entropy of the bubble and reservoir
is a functional of the profile
and a function of the other thermodynamic parameters,
\begin{eqnarray}
S_\mathrm{tot}[z]
& = &
S_\mathrm{b}(N,V,T)
- \frac{\gamma}{T} A[z]
- \frac{p}{T} V[z]
+ \frac{\mu}{T} N .
\end{eqnarray}
The solid-air and solid-liquid surface energies
(for both the substrate and the tip)
are constant because of the fixed contact radii
and so are neglected here.
Here $\mu = k_\mathrm{B}T\ln [ s p \Lambda^3 /k_\mathrm{B}T ]$
is the chemical potential of the air.
This expression is appropriate for the case that the bubble can exchange
number, volume, and area with the reservoir.

It is assumed that thermal equilibrium holds.
The entropy of the bubble may be taken to be that of an ideal gas
for thermal equilibrium,
\begin{equation}
S_\mathrm{b}(N,V,T) =
k_\mathrm{B} N \left[ 1 - \ln \frac{N \Lambda^3}{V} \right],
\end{equation}
where $\Lambda$ is the thermal wave length.
One could in fact use the entropy of a real gas instead of this
since the only things that enter below are its thermodynamic derivatives,
$p_\mathrm{b} = T \partial S_\mathrm{b}(N,V,T)/\partial V$,
and
$\mu_\mathrm{b} = - T\partial S_\mathrm{b}(N,V,T)/\partial N$.

Obviously, setting the number derivative of the total entropy
to zero yields the equilibrium condition
$ \overline \mu_\mathrm{b} = \mu$,
which for an ideal gas is the same as
\begin{equation}
\overline N = \frac{s p}{k_\mathrm{B}T} V[z] .
\end{equation}
Invoking the usual variational properties of equilibrium thermodynamics,
\cite{Attard02}
the number is fixed at this value in all that follows.

The functional derivative of the total entropy
with respect to the bubble profile is
\begin{equation}
\frac{\delta S_\mathrm{tot}[z]}{\delta z(r)}
=
\frac{\Delta_p}{T} \frac{\delta V[z]}{\delta z(r)}
- \frac{\gamma}{T} \frac{\delta A[z]}{\delta z(r)} .
\end{equation}
where $\Delta_p \equiv p_\mathrm{b}-p = (s-1)p$
is the excess pressure of the bubble.
One has
\begin{equation}
\delta V[z]  =
2 \pi \int_{r_1}^{r_3} \mathrm{d}r \, r \delta z(r) ,
\end{equation}
from which it follows that
\begin{equation}
 \frac{\delta V[z]}{\delta z(r)}  =
2 \pi  r  .
\end{equation}
Also
\begin{eqnarray}
\delta A[z] &= &
 2 \pi \int_{r_1}^{r_3} \mathrm{d}r \, r \frac{ z'(r)}{ \sqrt{1 + z'(r)^2} }
\delta z'(r)
 \\ \nonumber & = &
- 2 \pi \int_{r_1}^{r_3} \mathrm{d}r \,
\frac{\mathrm{d}}{\mathrm{d} r}
\left[ r \frac{ z'(r)}{ \sqrt{1 + z'(r)^2} } \right]
\delta z(r),
\end{eqnarray}
following an integration by parts and the vanishing of the perturbation
at the boundaries.
From this one has
\begin{equation}
 \frac{\delta A[z]}{\delta z(r)}  =
- 2 \pi  \frac{\mathrm{d}}{\mathrm{d} r}
\left[ \frac{ r z'(r)}{ \sqrt{1 + z'(r)^2} } \right] .
\end{equation}
Inserting these into the functional derivative of the total entropy
and setting the latter to zero
gives a differential equation (the Eular-Lagrange equation)
for the optimum profile,
\begin{equation}
0 =
2 \pi r \Delta_p
+
 \frac{\mathrm{d}}{\mathrm{d} r}
\left[ \frac{ 2 \pi r z'(r)}{ \sqrt{1 + z'(r)^2} } \right] \gamma .
\end{equation}
Due to diffusive equilibrium,
the excess pressure is a constant, $\Delta_p  = (s-1)p$.

The first integral of this is
\begin{equation} \label{Eq:int1}
\mbox{const.} =
\pi r^2 \Delta_p
+
\frac{ 2 \pi r z'(r)}{ \sqrt{1 + z'(r)^2} } \gamma .
\end{equation}

\subsubsection{Force} \label{Sec:Force}

Including a spring attached to the cantilever
with spring constant $k_\mathrm{t}$,
the extended total entropy is
\begin{equation}
S_{\mathrm{tot},k} = S_{\mathrm{tot}}([z],r_1)
- \frac{k_\mathrm{t}}{2T} [ z_0 - z_\mathrm{c} ]^2 .
\end{equation}
The cantilever spring is placed so that it is unextended
(zero force) when the tip is just in contact
with the undeformed bubble.

The derivative of the extended total entropy
with respect to the tip position,
at constant tip contact radius $r_1 $,
and evaluated at the optimum profile $\overline z(r;r_1,z_0)$
is now required.

Above, in deriving the Eular-Lagrange equation for the profile,
the variation at the boundaries vanished,
$\delta z(r_1) = \delta z(r_3) = 0$.
In the present case, the variation at contact on the tip must be
$\delta z(r_1) = \Delta z_0$.
This means that one picks up an extra term
from the integration by parts of the variation in area,
\begin{eqnarray}
\delta A[z] &= &
 2 \pi \int_{r_1}^{r_3} \mathrm{d}r \, r \frac{ z'(r)}{ \sqrt{1 + z'(r)^2} }
\delta z'(r)
\\  & = &
- 2 \pi  r_1 \frac{ z'(r_1)}{ \sqrt{1 + z'(r_1)^2} } \Delta z_0
\nonumber \\ && \mbox{ }
- 2 \pi \int_{r_1}^{r_3} \mathrm{d}r \,
\frac{\mathrm{d}}{\mathrm{d} r}
\left[ r \frac{ z'(r)}{ \sqrt{1 + z'(r)^2} } \right] \delta z(r).
\nonumber
\end{eqnarray}
Accordingly
\begin{eqnarray}
T \left. \frac{\partial S_{\mathrm{tot},k} }{\partial z_0}
\right|_{\overline z(r)}
& = &
\left[  \Delta_p \frac{\delta V[z]}{\delta z(r)}
- \gamma \frac{\delta A[z]}{\delta z(r)}
\right]_{\overline z(r)}
\frac{\partial \overline z(r)}{\partial z_0 }
\nonumber \\ && \mbox{ }
+ \pi r_1^2 \Delta_p
+  \frac{ 2 \pi  r_1 z'(r_1)}{ \sqrt{1 + z'(r_1)^2} } \gamma
\nonumber \\ && \mbox{ }
- k_\mathrm{t} [ z_0 - z_\mathrm{c} ]
\nonumber \\ &= &
0 + F + F_\mathrm{t} .
\end{eqnarray}
The first term vanishes for the optimum profile.
The remaining term proportional to $\Delta_p$ arises
from the derivative of the constant contributions to the volume.
This and the remaining term proportional to $\gamma$
give the force exerted by the bubble on the tip,
\begin{equation}
F = \pi r_1^2 \Delta_p
+  \frac{ 2 \pi  r_1 z'(r_1)}{ \sqrt{1 + z'(r_1)^2} } \gamma .
\end{equation}
This is just the pressure difference
times the cross-section contact area
plus the vertical component of the surface tension force
times the contact perimeter.

The quantity
$F_\mathrm{t} \equiv - k_\mathrm{t} [ z_0 - z_\mathrm{c} ]$
is the force exerted by the tip on the bubble.
One sees that in the equilibrium or static case,
when the extended total entropy is a maximum,
the force due to the bubble is equal and opposite
to the force due to the tip,
$ F = - F_\mathrm{t}(\overline z_0) $.
A positive bubble force,
as sketched in Fig.~\ref{Fig:geom},
corresponds to a negative cantilever force (in a signed sense).
In practice, one often calls the cantilever force
the applied force, or the load.
A positive cantilever force (negative bubble force)
gives an extended rather than a flattened bubble.

Comparing bubble force with the first integral, Eq.~(\ref{Eq:int1}),
one sees that the integration constant is just the force
exerted by the bubble on the tip,
so that one has
\begin{equation} \label{Eq:ELDE}
F =
\pi r^2 \Delta_p
+
\frac{ 2 \pi r z'(r)}{ \sqrt{1 + z'(r)^2} } \gamma .
\end{equation}
Hence one has a differential equation for the profile
as a function of the force, $F$, the substrate contact radius $r_3$,
the tip  contact radius $r_1$,
and the tip contact height $z_1$.
Given the geometry of the tip
(eg.\ the half angle $\alpha$),
the location of the tip of the tip, $z_0$,
is determined by the latter two quantities,
if it is ever required.

In what follows an analytic expression for the profile will
be derived in the weak force limit,
The expansion is valid when
$|F |R_\mathrm{c}/2\pi\gamma(s) r_1^2 \ll 1$.

\subsubsection{Location of Dimple Rim}

The dimple rim $r_2$ is the maximum height of the bubble,
so that $z'(r_2) = 0$.
From the differential equation for the profile this yields
\begin{equation}
r_2 = \sqrt{\frac{F}{\pi \Delta_p}}
=  \sqrt{\frac{F R_\mathrm{c}}{2 \pi \gamma}} .
\end{equation}
Note that since $r_2 \ge r_1$,
this sets a lower limit on the repulsive force
that gives rise to a dimple.
Of course there is also no dimple for attractive forces (negative loads).
In either case, the dimple rim plays no further role in the analysis.

\subsubsection{Profile}

Rearranging equation (\ref{Eq:ELDE}) for the profile,
gives
\begin{equation}
[ F - \pi \Delta_p r^2 ]^2 [1 + z'(r)^2]
= (2\pi\gamma r)^2 z'(r)^2 ,
\end{equation}
or
\begin{eqnarray} \label{Eq:zpr}
z'(r)
& = &
\frac{\pm [ F - \pi \Delta_p r^2 ]  }{
\sqrt{(2\pi\gamma r)^2 - [ F - \pi \Delta_p r^2 ]^2  } } .
\end{eqnarray}
The positive root is the physical root.

One can see that a minimum and a maximum force
is defined for a given pinned tip contact line $r_1$
when the gradient of the profile becomes infinite,
$z'(r_1) = \pm \infty$.
The bubble will rupture when the applied load exceeds these limits.
From the profile equation one has
\begin{equation}
F_\mathrm{min} =
- 2\pi r_1 \gamma \left[ 1 - \frac{r_1}{R_\mathrm{c}} \right] ,
\end{equation}
and
\begin{equation}
F_\mathrm{max} =
2\pi r_1 \gamma \left[ 1 + \frac{r_1}{R_\mathrm{c}} \right] .
\end{equation}

In fact, since the gradient of the profile can't be infinite anywhere,
these two limits hold for any $r$ on the interval $[r_1,r_3]$.
Using $r_1$ gives the tightest upper bound because $r \ge r_1$.
But it can be the case that a tighter lower bound
can occur by taking $r$ inside the interval.
In particular, if $r_1 \le (R_\mathrm{c}/2) \le r_3$,
then the bubble will rupture if $F < -\pi R_\mathrm{c} \gamma /2$.
If $r_3 \le  R_\mathrm{c}/2$,
then the bubble will rupture if
$F < - 2\pi r_3 \gamma [ 1 - {r_3}/{R_\mathrm{c}} ]$.

For small loads,
$|F | \ll 2\pi\gamma(s) r_1^2 /R_\mathrm{c}$,
one can expand the profile equation to linear order in the force,
\begin{eqnarray}
z'(r)
& = &
\frac{  F - \pi \Delta_p r^2   }{
\sqrt{(2\pi\gamma r)^2 - (\pi \Delta_p r^2)^2
+ 2(\pi \Delta_p r^2)F + {\cal O}(F^2)  } }
\nonumber \\ & = &
\frac{  (F/\pi\Delta_p) -  r^2
}{r \sqrt{ R_\mathrm{c}^2  - r^2 + 2(F/\pi\Delta_p)    } }
\nonumber \\ & = &
\frac{ -  r }{ \sqrt{ R_\mathrm{c}^2  - r^2   } }
+ \frac{  F/\pi\Delta_p }{r \sqrt{ R_\mathrm{c}^2  - r^2 } }
+ \frac{ r F/\pi\Delta_p }{ [R_\mathrm{c}^2  - r^2 ]^{3/2} }
\nonumber \\ & & \mbox{ }
+  {\cal O}(F^2) .
\end{eqnarray}
Recall that $\Delta_p = 2 \gamma/R_\mathrm{c}$.
The first term gives the undeformed profile, Eq.~(\ref{Eq:zcr}),
and the remainder give the perturbation due to the force to linear order.
With $\varepsilon(r) \equiv z(r) - z_\mathrm{c}(r)$,
this gives the derivative of the perturbation to linear order,
\begin{equation} \label{Eq:eps'}
\varepsilon'(r)
=
\frac{ F R_\mathrm{c} /2 \pi\gamma }{r \sqrt{ R_\mathrm{c}^2  - r^2 } }
+ \frac{ r FR_\mathrm{c} /2 \pi\gamma  }{ [R_\mathrm{c}^2  - r^2 ]^{3/2} } .
\end{equation}

The integral of this is
\begin{eqnarray} \label{Eq:vareps}
\varepsilon(r)
& = &
\frac{F}{2 \pi\gamma }
\left\{
\frac{1}{2} \ln \frac{ 1-\sqrt{1 - x^2} }{ 1+\sqrt{1 - x^2} }
+\frac{1}{\sqrt{1-x^2}}
- C_3 \right\} ,
\nonumber \\ &&
\end{eqnarray}
where $x\equiv r/R_\mathrm{c}$.
The integration constant is
determined by the condition that $\varepsilon(r_3) = 0$,
\begin{eqnarray}
C_3 \equiv
\frac{1}{2} \ln \frac{ 1-\sqrt{1 - x_3^2} }{ 1+\sqrt{1 - x_3^2} }
+ \frac{1}{\sqrt{1-x_3^2}} .
\end{eqnarray}


\begin{figure}[t!]
\centerline{
\resizebox{8.5cm}{!}{ \includegraphics*{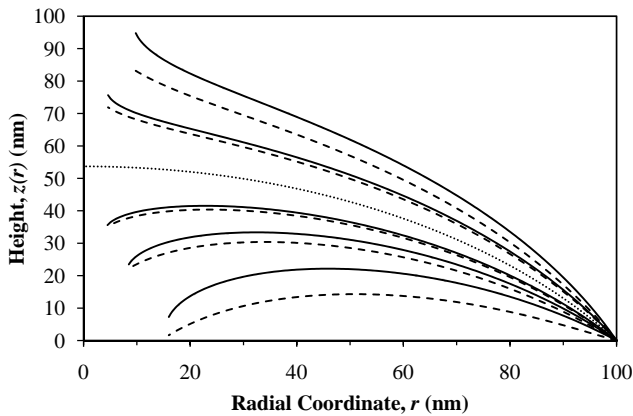} } }
\caption{\label{Fig:profile}
Pinned bubble profiles for various applied loads.
The dotted curve is the undeformed bubble (zero load),
a full curve is an exact profile,
and an adjacent dashed curve is the corresponding linear approximation.
Above and below the undeformed profile the loads are
$\pm 0.5\,$nN, $\pm 1\,$nN, and $+2\,$nN.
The parameters are $s=4$, $\gamma(s) = 0.018\,$N/m ($s^\ddag=5$),
and $r_3=100\,$nm, giving
$R_\mathrm{c}=120\,$nm, and $z_\mathrm{c}=54\,$nm.
Each curve terminates at its last stable contact radius.
}
\end{figure}

Figure \ref{Fig:profile} shows several profiles of deformed bubbles.
The exact profile was obtained by numerical integration of the
profile equation,  Eq.~(\ref{Eq:zpr}).
Here and below the linear approximation refers to analytic results based
the expansion to linear order in the force,
in this case Eq.~(\ref{Eq:vareps}).
In the case of the figure,  $2 \pi r_1^2 \gamma /R_\mathrm{c} = 0.4\,$nN
for $r_1 = 20$\,nm.
For loads with magnitude much less than this
the linear approximation can be guaranteed accurate accurate.
It can be seen in the figure that the performance
of the linear approximation is rather better than is indicated
by this parameter.
For larger loads there is a significant discrepancy
between the exact and the linear profile.
The problem is more acute for extensive than for compressive forces.

\subsubsection{Bubble Spring Constant}

The various quantities $z$ above were measured relative to the substrate.
Now, in the laboratory frame of reference,
let $\zeta_\mathrm{s}$ be the position of the solid substrate.
In this laboratory frame,
the position of the tip, the tip contact circle,
and the undeformed bubble interface are
\begin{equation}
\zeta_0 = z_0+\zeta_\mathrm{s},
\;\;
\zeta_1 = z_1+\zeta_\mathrm{s},
\mbox{ and }
\zeta_\mathrm{c} = z_\mathrm{c}+\zeta_\mathrm{s},
\end{equation}
respectively.
Initially, the substrate is at $\zeta_\mathrm{s} = 0$,
the tip is just touching the undeformed bubble
$\zeta_0 = z_\mathrm{c}$,
and the spring attached to the tip is undeflected.
Hence in general the deflection is
$\delta_\mathrm{t} = \zeta_0 - z_\mathrm{c}$,
and force exerted by the tip on the bubble is
\begin{equation}
F_\mathrm{t} = - k_\mathrm{t} \delta_\mathrm{t}
= -k_\mathrm{t} [ \zeta_0 - z_\mathrm{c}] ,
\end{equation}
where  $k_\mathrm{t}$ is the tip spring constant.

Now $z_1 = z_\mathrm{c}(r_1) + \varepsilon(r_1)$, or
\begin{equation}
\zeta_1 = \zeta_\mathrm{c}(r_1) + \varepsilon(r_1) .
\end{equation}
The amount of bubble deformation at contact is
\begin{equation}
\varepsilon(r_1) = -k_\mathrm{bub}^{-1} F,
\end{equation}
where $F$ is the force exerted by the bubble.
The bubble spring constant is given by the profile equation
evaluated at $r_1$,
\begin{equation} \label{Eq:kbm1}
k_\mathrm{bub}
=
-2 \pi\gamma
\left\{
\frac{1}{2} \ln \frac{ 1-\sqrt{1 - x_1^2} }{ 1+\sqrt{1 - x_1^2} }
+\frac{1}{\sqrt{1-x_1^2}}
- C_3 \right\}^{-1} .
\end{equation}
This depends upon the two pinning radii.
Obviously $R_\mathrm{c} > r_3 > r_1$.

\begin{figure}[t!]
\centerline{
\resizebox{8.5cm}{!}{ \includegraphics*{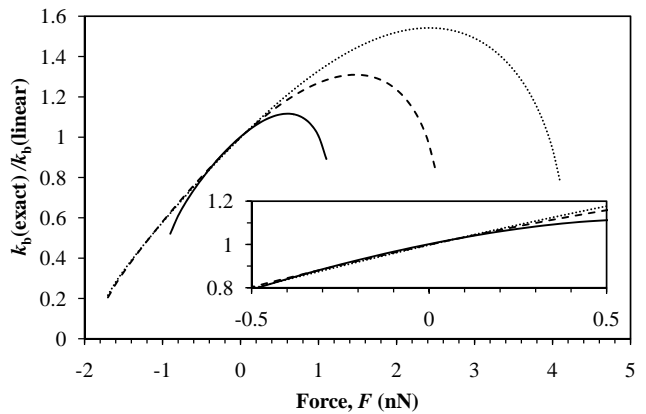} } }
\caption{\label{Fig:ratio-kb}
Ratio of exact effective bubble spring constant, $\Delta F/\Delta z(r_1)$,
to linear bubble spring constant $k_\mathrm{b}$, Eq.~(\ref{Eq:kbm1}),
as a function of the applied load.
The solid, dashed, and dotted curves are for
$r_1 =$ 10, 20, and 30$\,$nm, respectively.
Other parameters as in Fig.~\ref{Fig:profile}.
Each curve terminates at its limits of solution.
The inset magnifies the region around zero force.
}
\end{figure}

Figure \ref{Fig:ratio-kb} shows the exact
effective spring constant of the bubble,
$\Delta F/\Delta z(r_1)$ obtained from the numerical integration
of the profile equation,  Eq.~(\ref{Eq:zpr}),
normalized by the analytic expression obtained from the expansion
to linear order in the force,  Eq.~(\ref{Eq:kbm1}).
The expansion is valid when
$|F | \ll 2\pi\gamma(s) r_1^2 /R_\mathrm{c}$.
For the conditions in Fig.~\ref{Fig:ratio-kb},
the right hand side is 0.09, 0.38, and 0.85$\,$nN
for $r_1 =$ 10, 20, and 30$\,$nm, respectively.
One can indeed see in the figure
that the bubble behaves linearly to within about 20\% of the exact value
when the loads lie within this bound.

\begin{figure}[t!]
\centerline{
\resizebox{8.5cm}{!}{ \includegraphics*{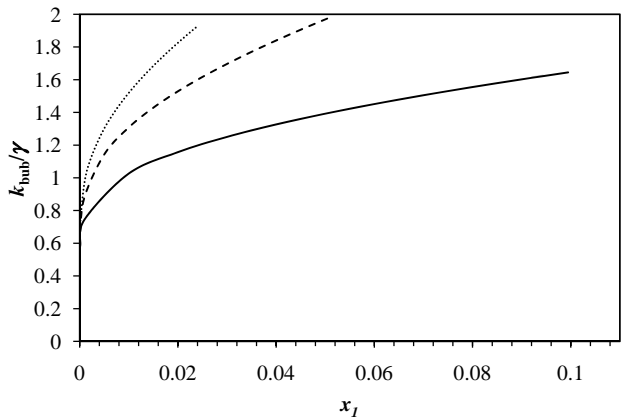} } }
\caption{\label{Fig:kb}
Bubble linear spring constant as a function of tip contact radius
$x_1 \equiv r_1/R_\mathrm{c}$
for a substrate radius $x_3 \equiv r_3/R_\mathrm{c}$
of 0.9 (solid curve), 0.7 (dashed curve), and 0.5 (dotted curve).
In each case, the upper limit is
$x_1^\mathrm{max} = [1 - \sqrt{1-x_3^2}]\tan \alpha$,
with $\alpha = 10^\circ$.
}
\end{figure}

Figure~\ref{Fig:kb} shows the ratio of the bubble linear spring constant
to the surface tension.
It can be seen that the variation is rather weak over the practical range,
with the bubble spring constant being not more than a factor of two larger
than the surface tension.
In general the bubble spring constant is larger than  the surface tension
unless the contact radius is very small.
Clearly, in the absence of specific information
about the size of the pinning radii,
one will not go too far wrong in taking
the liquid-vapor surface tension of the supersaturated interface
to be 0.5--1 times the measured bubble spring constant.


In the linear regime, the bubble and tip act as two springs in series.
To see this explicitly,
consider a change in the position of the substrate, $\Delta \zeta_\mathrm{s}$,
at constant $r_1$ (and $r_3$).
Since $\Delta r_1 = 0$,
the change in contact position must equal the change in tip position,
$\Delta \zeta_1 = \Delta \zeta_0 $.
In the static situation $\Delta F_\mathrm{t} =-\Delta F $,
and so one has
\begin{eqnarray}
k_\mathrm{t} \Delta \zeta_0
& = &
- k_\mathrm{bub} \Delta \varepsilon(r_1)
\nonumber \\ & = &
- k_\mathrm{bub} \left[ \Delta \zeta_1
- \Delta \zeta_\mathrm{c}(r_1)  \right]
\nonumber \\ & = &
- k_\mathrm{bub}  \left[ \Delta \zeta_0
- \Delta \zeta_\mathrm{s}  \right] .
\end{eqnarray}
This implies that
$ \left[  k_\mathrm{t} + k_\mathrm{bub} \right] \Delta \zeta_0
= k_\mathrm{bub} \Delta \zeta_\mathrm{s}$,
or
\begin{equation}
\frac{\Delta  \zeta_0}{\Delta \zeta_\mathrm{s}}
=
\frac{k_\mathrm{bub}}{k_\mathrm{t} + k_\mathrm{bub} }.
\end{equation}
This ratio is less than unity,
whereas for a hard surface it would be unity.
A measurement of the slope of the deflection versus the drive distance
allows the spring constant of the interface to be determined.
If the contact radii are known,
then the surface tension of the bubble can be estimated.

Alternatively,
in terms of
the separation between the tip and substrate,
$ z_0 \equiv \zeta_0 - \zeta_\mathrm{s} $,
one has
\begin{eqnarray}
\frac{\Delta  \zeta_0}{\Delta z_0}
& = &
\frac{\Delta  \zeta_0
}{\Delta  \zeta_0-\Delta  \zeta_\mathrm{s}}
\nonumber \\ & = &
\frac{1}{1-\Delta  \zeta_\mathrm{s}/\Delta  \zeta_0}
\nonumber \\ & = &
\frac{1}{ 1- \frac{k_\mathrm{t} + k_\mathrm{bub} }{k_\mathrm{bub}} }
\nonumber \\ & = &
\frac{-k_\mathrm{bub}}{ k_\mathrm{t} } .
\end{eqnarray}
In words, the slope of the deflection versus separation curve
equals the negative of the ratio of the bubble and cantilever
spring constants.

%
\section{Deformed Bubble: Prick Slip}
\setcounter{equation}{0} 
%

\begin{figure}[t!]
\centerline{
\resizebox{8.5cm}{!}{ \includegraphics*{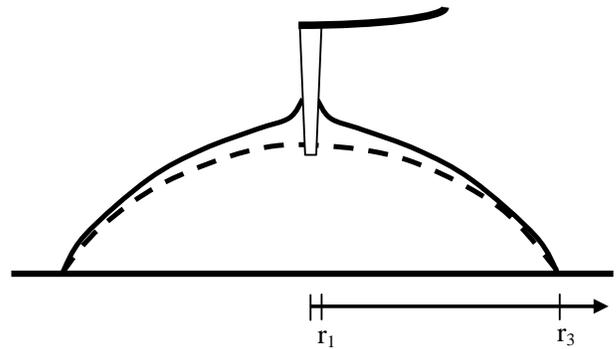} } }
\caption{\label{Fig:phobic}
Hemispherical bubble penetrated and deformed by a conical hydrophobic tip
(blunt radius $r_0$)
upon which it slips.
}
\end{figure}

For the case of slip on the tip,
the liquid-vapor-solid contact circle is not pinned at $r_1$
(see Fig.~\ref{Fig:phobic}).
One now has also to maximize the total entropy with respect to $r_1$,
taking into account the difference in solid surface energies,
$\Delta \gamma \equiv \gamma_\mathrm{sg} - \gamma_\mathrm{sl}$.
This is negative for a hydrophobic tip.

For a macroscopic bubble or droplet on a planar substrate
made of the same material as the tip,
the equilibrium condition is that
the contact angle measured in the liquid phase
satisfies $ \gamma \cos \theta_\mathrm{t} = \Delta \gamma $.
This does not hold in the present case
of a conical tip penetrating the bubble.

For the present case of slip,
the total entropy is
\begin{eqnarray}
S_\mathrm{tot}[z]
& = &
S_\mathrm{b}(N,V,T)
- \frac{\gamma}{T} A[z]
- \frac{p}{T} V[z]
+ \frac{\mu}{T} N
\nonumber \\ & & \mbox{ }
- \frac{\Delta \gamma}{T} A_\mathrm{t} .
\end{eqnarray}
The surface area of the blunt conical tip inside the bubble is
\begin{equation}
A_\mathrm{t} =
\frac{ \pi  r_1^2}{ \sin \alpha}
-\frac{ \pi  r_0^2}{ \sin \alpha}
+ \pi r_0^2.
\end{equation}
The formulae for the volume $V[z]$ and area $A[z]$ were given above
and depend upon $r_1$.
Also recall that $r_1-r_0 = [z_1-z_0] \tan \alpha$.

Differentiating with respect to the profile
(at constant $r_1$) gives the Eular-Lagrange equation
for the optimum profile, as given above.
Differentiating with respect to $r_1$,
one has $\mathrm {d} r_1 = \mathrm {d} z_1 \tan \alpha$,
which means that $\delta z(r_1) = \mathrm {d} r_1/\tan \alpha$,
and so the extra boundary term in the profile derivative
of the area appears, as in \S \ref{Sec:Force}.
Also, $ \mathrm {d} z_0 = 0$.
In view of this one has
\begin{eqnarray}
\lefteqn{
\left. \frac{\partial S_{\mathrm{tot}} }{\partial r_1}
\right|_{\overline z(r)}
}  \\
& = &
\left[  \frac{\Delta_p}{T} \frac{\delta V[z]}{\delta z(r)}
- \frac{\gamma}{T} \frac{\delta A[z]}{\delta z(r)}
\right]_{\overline z(r)}
\frac{\partial \overline z(r)}{\partial r_1 }
\nonumber \\ && \mbox{ }
+ \left[
\frac{2 \pi r_1^2}{\tan \alpha}
+ 2 \pi r_1 \left(z_0 - \frac{ r_0}{\tan \alpha} \right)
\right] \frac{\Delta_p}{T}
\nonumber \\ && \mbox{ }
+  \left[ 
2 \pi r_1 \sqrt{1 + z'(r_1)^2}
+ \frac{ 2 \pi  r_1 z'(r_1)}{ \sqrt{1 + z'(r_1)^2} \tan \alpha}
\right] \frac{\gamma}{T}
\nonumber \\ && \mbox{ }
- \frac{2 \pi r_1 \Delta \gamma}{T \sin \alpha}
\nonumber \\ &= &
\frac{ 2 \pi r_1 }{T}
\left\{ \left[
\frac{2 z_1}{R_\mathrm{c}}
+ \sqrt{1 + z'(r_1)^2}
+ \frac{ z'(r_1)/ \tan \alpha}{ \sqrt{1 + z'(r_1)^2} }
\right] \gamma
\right. \nonumber \\ && \left. \mbox{ }
- \frac{\Delta \gamma}{ \sin \alpha}
\rule{0cm}{0.6cm}\right\} .
\nonumber
\end{eqnarray}
Setting this to zero,
the trivial solution is $\overline r_1 = 0$.
The non-trivial solution gives an equation
that the profile slope at the optimum radius must satisfy,
\begin{equation} \label{Eq:CAC-cone}
\frac{\Delta \gamma}{ \gamma \sin \alpha}
=
\frac{2 z_1}{R_\mathrm{c}}
+\sqrt{1 + z'(r_1)^2}
+ \frac{ z'(r_1)/ \tan \alpha}{ \sqrt{1 + z'(r_1)^2} } .
\end{equation}
This replaces the planar contact angle condition.

In the limit $\alpha \rightarrow 0 $ this is
\begin{equation}
\frac{\Delta \gamma}{ \gamma }
=
\frac{ z'(r_1)}{ \sqrt{1 + z'(r_1)^2} }
=
\cos \theta_\mathrm{t} ,
\end{equation}
which is the expected contact angle condition
for a cylindrical tip.

The expression for the force ought to be unchanged.
Explicitly the force exerted by the bubble on the tip is
\begin{eqnarray}
F & = &
T
\frac{\mathrm{d} S_\mathrm{tot}}{\mathrm{d} z_0}
\nonumber \\ & = &
T \left\{
\int_{r_1}^{r_3} \mathrm{d}r \,
\left.
\frac{\delta S_\mathrm{tot}([z],r_1,z_0)}{\delta z(r)}
\right|_{[\overline z],\overline r_1}
\frac{\mathrm{d} \overline z(r)}{\mathrm{d} z_0}
\right. \nonumber \\ & & \mbox{ } \left.
+ \left.
\frac{\partial S_\mathrm{tot}([z],r_1,z_0)}{\partial r_1}
\right|_{[\overline z],\overline r_1}
\frac{\mathrm{d} \overline r_1}{\mathrm{d} z_0}
\right. \nonumber \\ & & \mbox{ } \left.
+ \left.
\frac{\partial S_\mathrm{tot}([z],r_1,z_0)}{\partial z_0}
\right|_{[\overline z],\overline r_1}
\right\}
\nonumber \\ & = &
T  \left.
\frac{\partial S_\mathrm{tot}([z],r_1,z_0)}{\partial z_0}
\right|_{[\overline z],\overline z_1}
\nonumber \\ & = &
\pi r_1^2 \Delta_p
+ \frac{2 \pi r_1 \gamma z'(r_1)}{\sqrt{1 + z'(r_1)^2} } .
\end{eqnarray}
Hence the expression for the profile is unchanged from the stick case,
although of course for a non-stick hydrophobic tip, $F< 0$.

\subsubsection{Algorithms}

Recall the equations of for the undeformed profile,
 Eq.~(\ref{Eq:zcr}) \emph{et seq.}
Also recall that in the linear approximation the deformation
$\varepsilon(r) \equiv z(r) - z_\mathrm{c}(r)$,
and its derivative $\varepsilon'(r) \equiv z'(r) - z_\mathrm{c}'(r)$,
are linearly proportional to the force,
Eqs~(\ref{Eq:eps'}) and (\ref{Eq:vareps}).
In view of these one can succinctly write
\begin{equation}
z(r_1) = z_\mathrm{c}(r_1) - k_\mathrm{bub}(r_1)^{-1} F ,
\end{equation}
and
\begin{equation}
z'(r_1) = z_\mathrm{c}'(r_1) + q_\mathrm{bub}(r_1) F .
\end{equation}
Write $k_1 \equiv k_\mathrm{bub}(r_1)$
and $q_1 \equiv q_\mathrm{bub}(r_1)$.

To linear order the optimum contact radius satisfies
(recall that $r_1-r_0 = [ z(r_1) - z_0 ] \tan \alpha$)
\begin{eqnarray}
0 & = &
\left[ \frac{2 z_1}{R_\mathrm{c}}
+
\sqrt{1 + z'(r_1)^2}
+ \frac{ z'(r_1)/ \tan \alpha}{ \sqrt{1 + z'(r_1)^2} }
\right] \gamma
- \frac{\Delta \gamma}{ \sin \alpha}
\nonumber \\ & = &
\left[
\frac{2 z_\mathrm{c}(r_1)}{R_\mathrm{c}}
+
\sqrt{1 + z'_\mathrm{c}(r_1)^2}
+ \frac{ z'_\mathrm{c}(r_1)/ \tan \alpha}{ \sqrt{1 + z'_\mathrm{c}(r_1)^2} }
\right] \gamma
- \frac{\Delta \gamma}{ \sin \alpha}
\nonumber \\ && \mbox{ }
- \frac{2 \gamma  F}{k_1 R_\mathrm{c}}
+ \left[
\frac{ z'_\mathrm{c}(r_1) \tan \alpha }{ \sqrt{1 + z'_\mathrm{c}(r_1)^2} }
+\frac{ 1}{ \sqrt{1 + z'_\mathrm{c}(r_1)^2} }
 \right. \nonumber \\ && \left.  \mbox{ }
-
\frac{z'_\mathrm{c}(r_1)^2
 }{ [{1 + z'_\mathrm{c}(r_1)^2}]^{3/2} }
\right]  \frac{\gamma q_1}{\tan \alpha} F
\end{eqnarray}
Rearranging this gives explicitly $F(r_1)$.

Let the equilibrium curve be
$\overline F(r_1)$, $\overline \zeta_\mathrm{s}(r_1)$.
The deflection is
$\delta_\mathrm{t} = z_0 + \zeta_\mathrm{s} - z_\mathrm{c}
= - F_\mathrm{t}/k_\mathrm{t} = F/k_\mathrm{t}$.
The equilibrium curve in linear approximation
can be generated as follows:
\begin{itemize}
\item
choose $r_1$

\item
calculate $F(r_1)$, and
$z_0 = z(r_1) + [r_0-r_1]/{\tan \alpha}$

\item
calculate $\zeta_\mathrm{s}$ from
$F(r_1) = -F_\mathrm{t}
= k_\mathrm{t} [ z_0+\zeta_\mathrm{s} - z_\mathrm{c}]$

\item
plot deflection, $\delta_\mathrm{t} = F(r_1) /k_\mathrm{t}$
versus separation, $ z_0 $.

\end{itemize}

For the exact, non-linear calculations,
one has to specify the load $F$
and calculate the profile $z(r;F)$ by numerical integration
from $(r_3, 0)$.
For the equilibrium (slip) case,
one terminates the profile at the value of $r_1$ where the profile
has the equilibrium contact angle, Eq.~(\ref{Eq:CAC-cone}).
In the cases where there are two solutions
one chooses the one based on continuity.
From the value of $z(r_1)$, one obtains $z_0$
and $\zeta_\mathrm{s}$  as in the linear case.
One then chooses a new load and repeats the process.
For the case of the pinned tip contact line,
for each $F$
one instead terminates the profile at the fixed value of $r_1$.


\subsubsection{Results}

\begin{figure}[t!]
\centerline{
\resizebox{8.5cm}{!}{ \includegraphics*{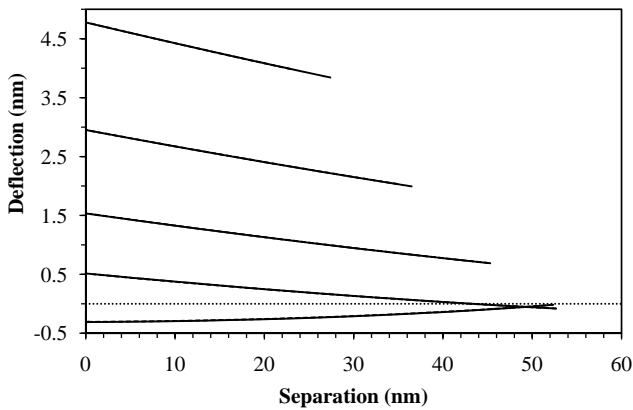} } }
\caption{\label{Fig:Equil-deflvsh-r0}
Equilibrium (slip) deflection versus separation curves
for a tip penetrating a nanobubble
for different tip radii.
These are the exact theory
for, from bottom to top at contact,
blunt tip radii of $r_0=$ 0, 20, 30, 40, and 50$\,$nm.
The dotted line is a guide to the eye.
The tip has $\alpha = 10^\circ$,  $k_\mathrm{t} = 0.35\,$N/m,
and $\Delta \gamma = 0\,$N/m or $\theta_\mathrm{t} = 90^\circ$.
All other parameters as in Fig.~\ref{Fig:profile}.
}
\end{figure}

Figure \ref{Fig:Equil-deflvsh-r0} shows several equilibrium
deflection versus separation curves.
Equilibrium here and below mean that the contact circle on the tip
is free to move. Hence the optimum contact angle that
maximizes the entropy is established at each separation.
For brevity this is also called slip.
The data in the figure is obtained with the exact theory.
Results obtained with the linear theory
are entirely obscured by the exact curves on the scale of the figure.

In the case of Fig.~\ref{Fig:Equil-deflvsh-r0},
the tip has been taken to be indifferent to water,
$\Delta \gamma = 0\,$N/m or $\theta_\mathrm{t} = 90^\circ$,
and for the most part the force is repulsive.
This corresponds to a positive cantilever deflection
and to a compressed bubble, as in Fig.~\ref{Fig:geom}
and in the lower half of Fig.~\ref{Fig:profile}.
The exception is for for the infinitely sharp tip, $r_0=0\,$nm,
which shows a weak attraction.

It is noticeable that the force curves are almost straight lines,
and that their magnitude increases with increasing tip radius.
The predominant reason that the force increases with decreasing separation
is the cone half angle,
which means that the contact radius increases as the tip penetrates further
into the bubble with decreasing separation.
The main reason that the force becomes more repulsive
with increasing end radius $r_0$
is that the repulsive pressure contribution is proportional to $r_1^2$
whereas the often attractive surface tension contribution
contains a factor $r_1$.

\begin{figure}[t!]
\centerline{
\resizebox{8.5cm}{!}{ \includegraphics*{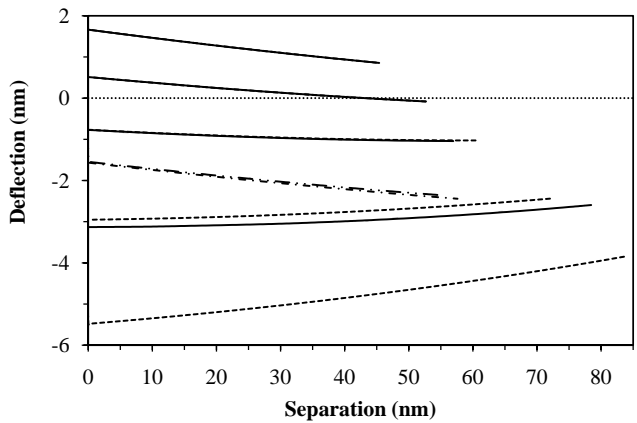} } }
\caption{\label{Fig:Equil-deflvsh-Dg}
Equilibrium (slip) deflection versus separation
for a blunt tip  penetrating a  nanobubble
for different tip surface energies.
The solid curves are the exact theory
and the dashed curves are the linear theory (obscured in the top three cases),
with $r_0=20\,$nm and, from bottom to top,
$\Delta \gamma = -10.0\,$mN/m ($\theta_\mathrm{t} = 98^\circ$),
$-6.3\,$mN/m ($\theta_\mathrm{t} = 95^\circ$),
$-2.5\,$mN/m ($\theta_\mathrm{t} = 92^\circ$),
$0\,$mN/m ($\theta_\mathrm{t} = 90^\circ$),
and
$+2.5\,$mN/m ($\theta_\mathrm{t} = 88^\circ$).
No exact solution was found for the lowest curve.
The dash-dotted curve and the partially obscured dash-double dotted curve are
the exact and linear results respectively
for  $r_0=50\,$nm,
and $\Delta \gamma =-6.3\,$mN/m ($\theta_\mathrm{t} = 95^\circ$).
All other parameters as in Fig.~\ref{Fig:Equil-deflvsh-r0}.
}
\end{figure}

Figure \ref{Fig:Equil-deflvsh-Dg} explores the effect of the tip surface energy
difference on the force curves.
The conversion of the surface energy difference to a macroscopic
tip contact angle uses the Young equation,
the saturation value of the surface tension, $\gamma^\dag = 0.072\,$N/m,
and assumes that the difference in surface energies is unchanged
by the level of supersaturation of the solution,
$\Delta \gamma = \gamma^\dag  \cos \theta_\mathrm{t}$.

It can be seen that as the surface energy difference becomes more negative,
the force at a given separation becomes increasingly attractive.
The penetrated bubble is extended from its undeformed shape
(see Fig.~\ref{Fig:phobic} and the upper half of Fig.~\ref{Fig:profile}).
The slope of the almost linear deflection versus separation curves
changes from negative to positive for
$\Delta \gamma \alt -5.0\,$mN/m ($\theta_\mathrm{t} \agt 92^\circ$),
close to where the surface energy difference changes sign,
and it becomes increasingly positive as the surface energy difference
becomes increasingly negative.

The linear theory becomes increasingly less accurate
as the surface energy difference becomes more negative.
Indeed, whilst the linear theory produced a solution curve
for $\Delta \gamma \le -10.0\,$mN/m at this $r_0=20\,$nm,
no exact solution was found (essentially because the bubble ruptured
at this contact radius).

Figure \ref{Fig:Equil-deflvsh-Dg} also presents the case of $r_0=50\,$nm
and $\Delta \gamma =-6.3\,$mN/m ($\theta_\mathrm{t} = 95^\circ$).
Compared to the same surface energy but for  $r_0=20\,$nm,
one sees that the force is less attractive
and as well the slope has decreased,
changing from positive to negative.
As mentioned above, increasing the contact radius
increases the repulsive pressure contribution to the force
more than the attractive surface tension contribution.
One can conclude that a negative slope is the signature
of pressure dominance
(large contact radius, small magnitude or positive surface energy difference)
whereas a positive slope is the signature of surface energy dominance,
(small contact radius, large negative surface energy difference).

In Figs~\ref{Fig:Equil-deflvsh-r0} and \ref{Fig:Equil-deflvsh-Dg}
it can be seen that there is a discontinuous jump from zero deflection
prior to contact with the undeformed bubble
to a non-zero deflection immediately after contact.
This is due to the fact
that non-contact forces are neglected in the present model
and also to the fact that the end of the tip has been taken to be a disc
(ie.\ perfectly blunt, planar).
In the present calculations liquid-vapor interface contact
with the flattened end has been excluded.
The size of this jump increases with the tip end radius $r_0$.
Obviously this is an idealized model of the actual tip of the tip,
which is actually neither perfectly sharp nor perfectly blunt,
and in practice there may be a smooth transition rather than a jump.
The cantilever manufacturer typically quotes
a tip  radius 20--60$\,$nm.
Also of course the AFM cantilever tip is generally
in the form of a square pyramid
rather than the present right circular cone.

\begin{figure}[t!]
\centerline{
\resizebox{8.5cm}{!}{ \includegraphics*{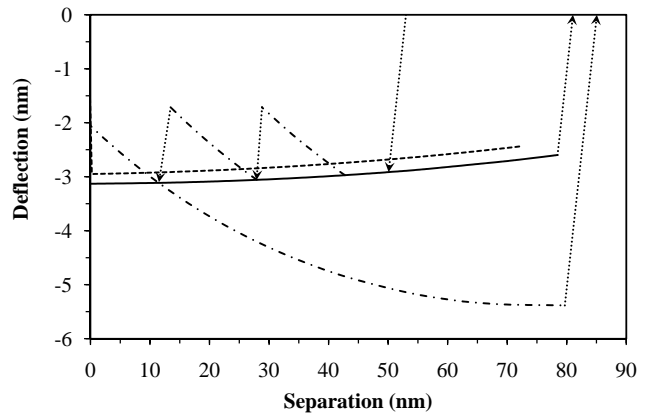} } }
\caption{\label{Fig:Stick-deflvsh}
Deflection versus separation
for a tip penetrating a nanobubble.
The solid curve is the equilibrium slip case using the exact theory,
and the parallel dashed curve uses the linear theory.
The dash-dotted curve is a stick-slip case using the exact theory,
with the pinned contact radii being $r_1=$ 26.4, 28.9, and 31.6$\,$nm
(chosen more or less arbitrarily).
The approach curve overlaps the initial part of the retraction curve
on the final stick region.
The dotted arrows show cantilever jumps.
The final jump out is at the limit of bridging bubble stability
for $r_1=  31.6\,$nm.
The blunt tip radius is  $r_0=20\,$nm,
and the difference in tip surface energies is
$\Delta \gamma = -6.3\,$mN/m ($\theta_\mathrm{t} = 95^\circ$),
with all other parameters Fig.~\ref{Fig:Equil-deflvsh-r0}.
}
\end{figure}

Figure \ref{Fig:Stick-deflvsh}
shows both a slip trajectory and
a stick-slip trajectory for a blunt tip, $r_0=20\,$nm.
The stick branches were chosen more or less randomly,
with one eye on aesthetics,
one eye on fundamental considerations,
and one eye on experimental data.
There is no fundamental reason that when the contact line gives way
it should jump to, and immediately stick at, the equilibrium position.
It could stick prior to reaching the equilibrium position,
or it could slip along the equilibrium curve after the jump.
One could perhaps argue for a yield stress
such that the contact line always slipped when the excess force
per unit contact line  reached a certain value,
but this has not been done here.

The first tip contact line stick has been taken to occur
at $r_1= 26.4\,$nm.
For the exact calculation,
the initial slope of the stuck branch is -0.078,
and the average slope of the entire branch is -0.090.
The linear prediction for the slope at this contact radius
and surface tension is -0.137.

The second tip contact line stick occurs at $r_1= 28.9\,$nm,
and the exact initial slope is -0.86,
and the average slope of the entire branch is -0.95.
The linear prediction for the slope at this contact radius is -0.143.

The third tip contact line stick occurs at $r_1= 31.6\,$nm,
and the exact initial slope of the stuck branch is -0.083,
and the average slope (extension, into contact)  is -0.093.
The linear prediction for the slope at this contact radius is -0.150.

For the single stuck branch on retraction,  $r_1= 31.6\,$nm,
curvature is quite evident,
which graphically indicates that the linear theory is inapplicable.
This curvature, including the final flat region,
is consistent with the data for the effective exact bubble spring constant
in Fig.~\ref{Fig:ratio-kb}.
The vanishing slope is due to the fact that
the  bubble spring constant tends to zero with increasing bubble extension.

It is evident that in the case treated in Fig.~\ref{Fig:Stick-deflvsh},
the ratio of the exact slope to the linear slope is about 2:3.
This is in agreement with the ratio indicated in Fig.~\ref{Fig:ratio-kb}
for $F=-1\,$nN.
Fitting the linear approximation
to measured experimental data where stick is evident,
particularly on the extension branch close to the equilibrium curve,
can be expected to provide
a useful first estimate for the bubble surface tension.


%
\section{Experiment}
\setcounter{equation}{0}  \setcounter{subsubsection}{0}
%

\subsection{Limitations of Theory}

Before analyzing any experimental data,
it is worth enumerating the limitations of the present theory.

First, the theory models the tip as a right circular cone,
whereas in reality an AFM tip is a rectangular pyramid.
Second, it models the tip of the tip as perfectly blunt,
a disc of radius $r_0$,
whereas in reality the tip may be curved from wear.
Third, it is assumed that the cross-section of the nanobubble
is perfectly circular,
which need not be the case if substrate heterogeneity determines
where the contact line is pinned.
Fourth, it is assumed that the substrate contact line does not alter during
the penetration of the nanobubble by the tip.

Fifth, it is assumed that the tip penetrates the nanobubble at the apex,
whereas in reality it can penetrate off-axis, either by design or by accident.
Sixth, it is assumed that the tip is oriented normal to the substrate,
whereas in reality the cantilever and tip are tilted at about 11$^\circ$
in the axial plane.
Seventh,
it is assumed that the air inside the nanobubble remains
in diffusive equilibrium with that in the solution
during the force measurement (constant chemical potential),
whereas in reality the measurement might be rapid enough
for constant number to hold instead.

The fifth and sixth points mean that
the predicted  normal force on the cantilever has an error
that increases with displacement from the central axis of the nanobubble,
that the displacement from the central axis varies with the separation
and with the deflection
of the cantilever during a force measurement (due to the tilt),
and that there is a torque on the cantilever due to the asymmetric
forces on the tip when it passes
through the nanobubble interface off-axis.
This nanobubble torque is in addition to the torque
that acts in all AFM force measurements on the tip
in contact with the hard substrate
due to the normal and lateral (friction) forces,
which are implicitly included in the photo-diode calibration.
\cite{Stiernstedt05,Attard06,Attard13}

The limitation summarized in the fifth point can be substantially alleviated
by ensuring that the separation at which first contact
with the nanobubble occurs is equal to
the measured height of the undeformed nanobubble.
One is aided in this by the fact that the nanobubble profile
is horizontal near the apex.
In this case one can be confident that,
at least for small cantilever deflections
and separations close to initial contact,
the tip is penetrating the nanobubble close to the apex,
and the calibration factor is correct.
In this regime the present calculation of the nanobubble normal force
and the neglect of any nanobubble lateral forces or torques,
ought to be accurate.
The consequences of the seventh point are discussed below
on p.~\pageref{para:const-N}.

In view of these limitations,
one ought not to expect the present theory to be able to quantitatively
describe every aspect of an individual nanobubble force measurement.
The main goal in the first instance is to establish
that the nanobubble surface tension is less than the usual air-water
surface tension, and if possible to quantify reliably its value
for a given solution.

To this end the following protocol was adopted.
Primary emphasis was placed on the first pinned region
of the measured separation-deflection curve,
since this is the one that can be guaranteed  closest to the apex.
Additional pinned regions were used to confirm
the values deduced from the first one.

\comment{ 
From the linear fit to a pinned region,
the slope $\chi$ and the separation at which
the pinned region would extrapolate to zero deflection, $z_\mathrm{t0}$,
were obtained,
\begin{equation}
\delta_\mathrm{t} = \chi [\zeta_0-\zeta_\mathrm{c} ],
\end{equation}
where $\delta_\mathrm{t}$ is the cantilever deflection
(the force is $F = k_\mathrm{t} \delta_\mathrm{t} $)
and $z_\mathrm{t}$ is the separation.

The radius of the pinned tip contact line for this particular branch,
$r_1$, can be determined as follows.
At the extrapolated separation of zero force, $z_\mathrm{t0}$,
by definition the nanobubble must be undeformed,
and so at this separation
the height of the contact line above the substrate must satisfy
\begin{equation}
z_1 = z_\mathrm{c}(r_1)
\equiv z_\mathrm{c} - R_\mathrm{c} + \sqrt{R_\mathrm{c}^2- r_1^2} .
\end{equation}
From the conical geometry one must have
\begin{equation}
r_1 - r_0 = [ z_1 - z_\mathrm{t0}] \tan \alpha .
\end{equation}
To leading order $z_1 = z_\mathrm{c} + {\cal O}(r_1^2/R_\mathrm{c})$,
which is a good starting point for an iterative solution
of these two equations.
One can solve these two equations for $z_1$ and $r_1$
in terms of the unknown radius of the blunt tip $r_0$.
One can now repeat this process for each of the pinned regions evident
in the data.
Since the surface tension can be determined from the slope $\chi$
and $r_1$ (see below),
the most consistent value of $r_0$
is the one that yields a single value of the surface tension
for all the branches.

} 

Since the surface tension that is required to fit a given slope
decreases with increasing value of  $r_1 \agt r_0  $,
one can establish an upper bound for the surface tension
by specifying the lowest realistic value of $r_0$.
In view of the specification that a new tip has radius
in the range 20--60~nm,
fixing for example  $r_0=20\,$nm should give an upper limit
on the surface tension.

Further, as is discussed below on p.~\pageref{para:const-N},
for a given surface tension and pinned contact radius $r_1$,
the slope calculated at constant chemical potential
is less in magnitude than the slope calculated at constant number.
This means that the surface tensions obtained below at constant
chemical potential are larger than those that would be required
to fit the slopes at constant number.
Again one can be confident that the surface tensions obtained here
are an upper bound on those for actual nanobubbles.

Since the slope of the pinned regions equals the
negative of the ratio of the nanobubble  spring constant
to the cantilever spring constant,
the surface tension can now be determined
using the linear theory.
The accuracy of this can be checked against the exact theory.
From the surface tension and the nanobubble curvature radius
determined by tapping mode imaging,
the supersaturation ratio is now determined, Eq.~(\ref{Eq:Rcrit}).
Using a linear model for the supersaturated surface tension,
$\gamma(s) =(s^\ddag - s) \gamma^\dag /(s^\ddag-1)$,
\cite{Moody03,Moody04,He05}
where $\gamma^\dag=0.072\,$N/m is the saturated surface tension,
the spinodal supersaturation ratio $s^\ddag$ can now be determined.

\subsection{Nanobubble 1}

\begin{figure} 
\centerline{\resizebox{8.5cm}{!}{ \includegraphics*{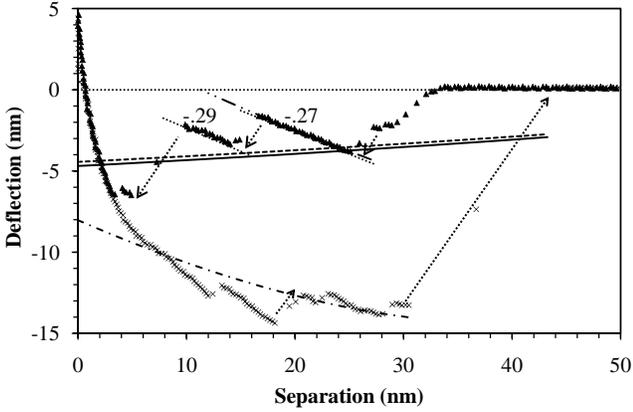} } }
\caption{\label{Fig:AFM-F3}
Cantilever deflection versus separation
for a SiN tip penetrating a nanobubble on an HOPG substrate
on approach (triangles) and retraction (crosses).
The cantilever has
spring constant $k_\mathrm{t}=0.35\,$N/m
and conical half-angle $\alpha = 10^\circ\!$.
The undeformed nanobubble has measured
height $z_\mathrm{c}=33\,$nm
and substrate contact radius $r_3 =108\,$nm,
corresponding to a curvature radius $R_\mathrm{c} = 192\,$nm
and a contact angle of $146^\circ$.
The dotted lines and arrows are guides to the eye,
with the adjacent number giving the slope.
The full and dashed curves are respectively the calculated
exact and linear equilibrium deflection
(tip contact line slip) with $r_0=10\,$nm,
$\gamma = 0.040\,$N/m,
$s=5.16$, 
and,
$\Delta \gamma=-10.0\,$mN/m (equivalently, $\theta_\mathrm{t}=98^\circ$).
The dash-double dotted line ($r_1=10\,$nm, $r_0=6\,$nm, obscured),
and the dash-dotted curve ($r_1=20\,$nm, $r_0=10\,$nm),
are calculated exact deflections with the tip contact line pinned,
using $\gamma = 0.044\,$N/m, $s=5.55$. 
}
\end{figure}

Figure \ref{Fig:AFM-F3}
shows AFM measurements of the force on a cantilever tip
due to a single nanobubble.
What is plotted is the positional deflection of the cantilever;
to obtain the force, multiply by the cantilever spring constant,
$k_\mathrm{t}=0.35\,$N/m.
The first measurably significant deflection
occurs at a separation of $z_0 = 32.9\,$nm.
This is in good agreement with the height
of this particular nanobubble imaged in tapping mode, $z_\mathrm{c}=33\,$nm.
This suggests that pre-contact forces
(van der Waals, electric double layer)
are negligible.
It also confirms that the measurement was performed
in the central region of the nanobubble close to the apex.

The nanobubble profile obtained from the image (not shown)
has a contact radius of $r_3 =108\,$nm,
which, with its height,
corresponds to a contact angle of $146^\circ$.
The fact that this contact angle is substantially higher
than the contact angle of a macroscopic water drop on HOPG,
64--92$^\circ$,\cite{Shin11,Li13}
is strong evidence that the nanobubble contact rim is pinned.\cite{Attard15}

In the experimental data just after first nanobubble contact,
the positively sloped region, followed by a brief plateau,
followed by a small jump to the base of the first marked linear region,
are all due to the initial spreading of the nanobubble
on the tip of the tip and up its sides.
This region is not well-modeled by the present geometry
of the tip of the tip as a perfectly planar circular disc.
The fact that the deflection is negative in this region
indicates that it is favorable for the tip to penetrate the nanobubble,
which is to say that the SiNi tip must be hydrophobic,
or possibly barely hydrophilic.

The two linear regions with labeled slopes
evident in the experimental deflection data on approach
confirm that the bubble can have a pinned contact line
and behave as a Hookean spring.
Since the cone half angle is small, $\alpha = 10^\circ\!$,
to leading order one can take the contact radius used
in fitting the slopes to be the same as the
radius of the perfectly blunt tip, $r_1 \approx r_0$.
The manufacturer's quoted radius of the tip,
20--60$\,$nm,
which might refer to either the tip's width
or else its radius of curvature,
can be assumed to be of the same order as $r_0$
in the present simple model.

Choosing a contact radius at the lower end of the realistic values,
$r_1 = 10\,$nm,
a value of $\gamma = 0.040\,$N/m gives
$k_\mathrm{b}/k_\mathrm{t} = 0.27$,
which is the negative of the measured slope of the first linear region.
Using Eq.~(\ref{Eq:Rcrit}),
this surface tension requires a supersaturation ratio of
$s=5.16$ to give a critical radius of $R_\mathrm{c}=193\,$nm,
equal to that deduced from the tapping mode
images of this particular nanobubble.
Conversely, choosing the upper limit $r_1 = 50\,$nm,
the fitted surface tension is $\gamma = 0.015\,$N/m and $s=2.58$.
Hence even in the most pessimistic case of smallest contact radius
one can see that the nanobubble surface tension,  $\gamma = 0.040\,$N/m
almost a factor of two smaller than
the  surface tension of the saturated air-water interface,
 $\gamma = 0.072\,$N/m,
and that the solution is substantially supersaturated with air, $s=5$.

The just quoted slopes were obtained with the linear theory.
Applying the exact non-linear theory
with the contact line again pinned at $r_1=10\,$nm,
the tangent at the start of the dash-double dotted curve
in Fig.~\ref{Fig:AFM-F3},
gives the required slope $-0.27$ using $\gamma = 0.044\,$N/m
and $s=5.55$.
In this case using $r_0=6\,$nm shifts the curve laterally
to coincide with the measured data.
Obviously using larger values of $r_1$ will require smaller
values of $\gamma(s)$ to fit the slope.
There is very little curvature evident in the non-linear curve.
The 10\% difference in the surface tension
between the exact and the linear fits
means that the linear theory provides an acceptable
estimate of the surface tension from the slope
that is both analytic and reliable.

The slope of the second linear region in  Fig.~\ref{Fig:AFM-F3}, $-0.29$,
is fitted by  $\gamma = 0.043\,$N/m using $r_1 = 10\,$nm
using the linear theory.
Alternatively,
the change in contact position of the nanobubble on the tip
may be approximated as
the change in separation at the base of the two linear regions,
$\Delta z_1 \approx \Delta z_0 = 11\,$nm,
assuming that the bubble profile is essentially the same
in the two cases,
which it would be if the contact line were mobile
prior to the start of the pinned regions.
The change in contact radius is
$\Delta r_1 = \Delta z_1 \tan \alpha \approx 2\,$nm.
Using this one finds that a single surface tension
(to better than 0.03\%)
$\gamma = 0.0402\,$N/m
gives a slope of -0.27 using $r_1 = 10\,$nm,
and a slope of -0.29 using $r_1 = 12\,$nm.
In the case of $r_1 = 50\,$nm and $r_1 = 52\,$nm,
the two slopes are given by a single surface tension
with a slightly worse variance of $2\%$.
This tends to suggest that the smaller contact radius is more applicable,
but this is by no means conclusive.
In any case, that a single surface tension combined with the geometry
of the tip fits the two slopes supports the model
and the value of the surface tension.

\label{para:const-N}
Here and below the  calculations are performed
at constant chemical potential,
which is computationally convenient.
Although there is strong evidence (see below)
that the nanobubble is in diffusive equilibrium with the solution
over the time of the series of force measurements,
it is unclear whether it is best to model each force measurement
as at constant chemical potential or as at constant number.
For the purposes of comparison,
some exact calculations have been carried out at constant number.
Using a surface tension of $\gamma(s) = 0.044\,$N/m,
at a pinned radius of $r_1=10\,$nm
the tangent at zero force at constant chemical potential is -0.30,
compared to  -0.41 at constant number.
Using instead  $r_1=20\,$nm and the same surface tension,
the tangent at zero force at constant chemical potential is -0.40,
compared to -0.62 at constant number.
One sees that the slope has a higher magnitude at constant number,
and that it increases relatively more rapidly with contact radius.
Hence one would require smaller surface tensions to fit the measured slopes
if one used constant number.
The calculations here and below are at constant chemical potential,
and so the surface tensions obtained represent an upper bound
on the actual nanobubble surface tension.

The two almost horizontal curves in  Fig.~\ref{Fig:AFM-F3}
are the calculated exact and linear equilibrium curves,
which assume that the contact line is mobile on the tip.
The good agreement between the linear and the exact calculations
is somewhat better than
that for the predicted effective bubble spring constant.
Both equilibrium calculations use
$r_1 = 10\,$nm, $\gamma = 0.04\,$N/m, and $s=5.16$.
In addition a surface energy difference of
$\Delta \gamma=-10.0\,$mN/m was fitted,
which corresponds to a macroscopic contact
of water on a planar SiNi substrate of  $\theta_\mathrm{t}=98^\circ$.
This is slightly hydrophobic.
The criterion for the fit, which was done by eye, was that the curve
should pass close to the base of the two linear regions.
(For reasons that are discussed below,
the end point of the final jump was not included in the fit.)
Since the cantilever jumps to these bases,
the nanobubble at contact must also be moving along the tip,
and so it can probably be assumed that the contact position
is the equilibrium one.
Of course the contact line may become pinned at the end of the jump
prior to achieving its equilibrium position,
so this fit may underestimate the magnitude of the surface energy difference.
Also, using a larger contact radius would require a smaller in magnitude value
for the fit.
Fortunately,
the surface tension obtained by fitting the slopes of the linear regions
is not affected by the  tip solid surface energies.

From the fitted value of the surface tension at $r_1=10\,$nm,
$\gamma = 0.040\,$N/m,
and the measured nanobubble curvature radius,
$R_\mathrm{c} = 193\,$nm,
which is equal to the critical radius, Eq.~(\ref{Eq:Rcrit}),
the supersaturation ratio can be deduced to be $s=5.2$.
The linear model for the supersaturated surface tension is
$\gamma(s) =(s^\ddag - s) \gamma^\dag /(s^\ddag-1)$,
where $\gamma^\dag=0.072\,$N/m is the saturated surface tension,
and $s^\ddag$ is the spinodal saturation ratio.
This has been shown to fit the available computer simulation data
reasonably accurately.\cite{Moody03,Moody04,He05}
These computer simulations give $s^\ddag \approx $4--6
for a Lennard-Jones fluid, depending on the temperature.
The present fit, $\gamma(5.2) = 0.040\,$N/m,
gives the spinodal supersaturation ratio $s^\ddag = 10.4$.
Alternatively, at $r_1=50\,$nm,
the linearly fitted supersaturated surface tension
$\gamma=0.015\,$N/m requires $s=2.6$
to give the measured nanobubble curvature radius,
and corresponds to $s^\ddag=3.0$ in the linear model.

Also shown in Fig.~\ref{Fig:AFM-F3}
is the calculated exact (ie.\ non-linear) deflection on extension
with the contact radius pinned at $r_1=20\,$nm.
The value of the contact radius was chosen so that the calculated
curve fitted by eye the measured data at the end of the retraction branch.
This calculation used the non-linear fitted value
$\gamma = 0.044\,$N/m and $s=5.55$,
and also $r_0=10.3\,$nm, which shifts the curve laterally.
The consistency of this with the exact fit to the slope $-0.27$
($r_1=10\,$nm and $r_0=6.2\,$nm)
is probably already acceptable;
with a little optimization of $r_1$ it could doubtless be made even better.
It is also undoubtable that larger values of $r_1$ and $r_0$
and smaller values of $\gamma(s)$ could also fit
both the slope of the pinned regions on approach
and the flattened region on retraction.
The conclusion that one can draw is that for separations
$z_0 \agt 20\,$nm the retraction data can be described
by the pinned nanobubble model using parameters consistent
with what was deduced from the slopes of the pinned approach data.

\label{Para:Hook}
For separations $z_0 \alt 20\,$nm on retraction,
and $z_0 \alt 3\,$nm on approach,
the data in Fig.~\ref{Fig:AFM-F3}
is not described by either the pinned or the mobile
contact line theory.
Similar steep curved regions
have been observed in a number of other AFM nanobubble measurements,
albeit for colloid probes rather for tips.
\cite{Carambassis98,Tyrrell02}
The origin of this particular behavior is unclear.
Because of the coincidence of approach and retraction here,
this is clearly an equilibrium, non-dissipative phenomenon.
Calculations show that it is not due to elastic
deformation of the substrate (not shown).\cite{Attard92,Attard01b}
It might be due to torque on the cantilever,
although measurements across the nanobubble indicate
that this is in general negligible.
That the curve is much steeper at $F=0$ than the clearly pinned regions
suggests that it is not due to pinning of the contact line,
unless the pinned contact radius had increased very substantially.
The apparently contiguous flat region $r\agt 20\,$nm on retraction
is similar to the non-linear calculations
of the force due to pinning of a highly extended nanobubble
with small contact radius.
Obviously whatever the origin of this behavior,
it could be simply additive to the force due to the pinned
(or slipping) contact line since the nanobubble force
is always present.
Because of the uncertainty as to the origin of this force close to contact
it has been neglected in fitting  the nanobubble.

The measurements in Fig.~\ref{Fig:AFM-F3}
were part of a sequence of twelve successive force measurements
across this particular  nanobubble (not shown).
The number, position, and extent of the linear regions
could differ between force measurements,
presumably because contact line stick and slip are stochastic events,
but the slopes were unchanged.
This suggests that torque on the cantilever due to off-apex penetration
has negligible effect.
Tapping mode images before and after the sequence of force measurements
show that the nanobubble itself
was unchanged in size and shape by the force measurements.
This is strong evidence that the nanobubble is thermodynamically stable,
and that even penetrating it a dozen times with the cantilever tip
did not destroy or alter it.
It is also evidence that the nanobubble is pinned at it contact rim.
In several other series of measurements,
up to a hundred force measurements were performed on a single nanobubble,
interspersed with several AFM tapping mode images,
and no significant change in the nanobubble was observed in any case.


\subsection{Nanobubble 2}

\begin{figure}[t!]
\centerline{\resizebox{8.5cm}{!}{ \includegraphics*{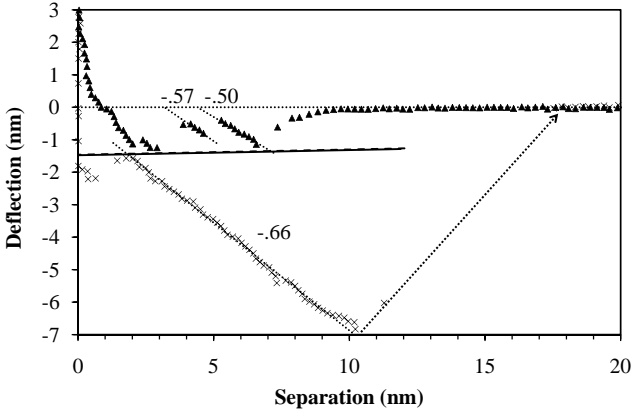} } }
\caption{\label{Fig:AFM-F7}
Measured cantilever deflection versus separation
for a nanobubble
(curves and lines as in preceding figure;
different  nanobubble, cantilever, and solution).
The cantilever has
spring constant $k_\mathrm{t}=0.24\,$N/m,
and conical half-angle $\alpha = 10^\circ\!$.
The undeformed nanobubble has measured
height $z_\mathrm{c}=9.6\,$nm
and substrate contact radius $r_3 =82.5\,$nm,
corresponding to a curvature radius $R_\mathrm{c} = 359\,$nm
and a contact angle of $167^\circ$.
The calculated equilibrium exact (solid) and linear (dashed, obscured) curve
use $r_0 = 10\,$nm,
$\gamma = 0.041\,$N/m,
$s=3.3$, 
and $\Delta \gamma=+1.3\,$mN/m (equivalently, $\theta_\mathrm{t}=89^\circ$).
}
\end{figure}

Figure \ref{Fig:AFM-F7} shows results for another nanobubble
in a different solution,
with stick-slip behavior evident.
Assuming an initial tip contact radius of $r_1=10\,$nm,
the measured slope $-0.50$ corresponds to a surface tension of
$\gamma(s) = 0.041\,$N/m (linear approximation).
Using instead $r_1=20\,$nm gives $\gamma(s) = 0.028\,$N/m.
Larger contact radii require even smaller surface tension to yield this slope.
Exact calculations differ by less than 1\% from the linear results
for the bubble spring constant in this regime.

The second pinned region with slope -0.57
corresponds to $\gamma(s) = 0.047\,$N/m using $r_1=10\,$nm,
and to $\gamma(s) = 0.032\,$N/m using $r_1=20\,$nm.
The third pinned region with slope -0.66
corresponds to $\gamma(s) = 0.054\,$N/m using $r_1=10\,$nm,
and to $\gamma(s) = 0.037\,$N/m using $r_1=20\,$nm.

From the point at which the pinned regions extrapolate to zero deflection,
one can deduce the value of the pinned radius $r_1$
for a specified value of the tip radius $r_0$.
However, it is not possible to find a single value of $r_0$
which yields a single surface tension when all three slopes are fitted.
For example, fixing $r_0 = 10\,$nm,
one finds that the three slopes -0.50, -0.57, and -0.66
correspond to $r_1 =$ 10.9, 11.1, and 11.7$\,$nm,
and to $\gamma(s) =$ 0.039, 0.045, and 0.050$\,$N/m, respectively.
There is nothing wrong with $r_1$ increasing
with each successive pinning event as the tip penetrates the nanobubble,
but one would have hoped for a single surface tension.
The best that can be done (ie.\ minimizing the sum of the relative
standard deviations in surface tension and in tip radius),
yields $r_0 = 12.323 \pm .001\,$nm and
$\gamma(s) = $0.036,  0.041, and 0.046$\,$N/m, respectively.
Possibly doing the calculations at constant number
rather than the present constant chemical potential
might yield more consistent results (see p.~\pageref{para:const-N}).
Or possibly the problem is related to the unknown origin
of the steep hook at small separations discussed on p.~\pageref{Para:Hook}.
In any case, the preferred value of surface tension
is the one taken  from the very first pinned region,
since this lies closest to the tip of the tip of the cantilever
and to zero force.

The surface tension obtained using $r_1=10\,$nm,
$\gamma(s) = 0.041\,$N/m,
and the nanobubble radius of curvature  $R_\mathrm{c} = 359\,$nm
correspond to a supersaturation value of $s=3.3$.
Using this in the linear model for the supersaturated surface tension
gives a spinodal supersaturation ratio of $s^\ddag =6.3$.
Using instead $r_1=20\,$nm, $\gamma(s) = 0.028\,$N/m,
correspond to a supersaturation value of $s=2.6$.
and a spinodal supersaturation ratio of $s^\ddag =3.6$.

Figure \ref{Fig:AFM-F7} also shows equilibrium calculations
using  $r_0=10\,$nm, $\gamma(s) = 0.041\,$N/m, and  $s=3.3$.
The value of the surface energy difference,
$\Delta \gamma=+1.3\,$mN/m (equivalently, $\theta_\mathrm{t}=89^\circ$)
was chosen so that the curves passed through the base
of the first pinned region.
The exact and the linear calculations
are almost indistinguishable.


\subsection{Nanobubbles 3 and 4}

\begin{figure}[t!]
\centerline{\resizebox{8.5cm}{!}{ \includegraphics*{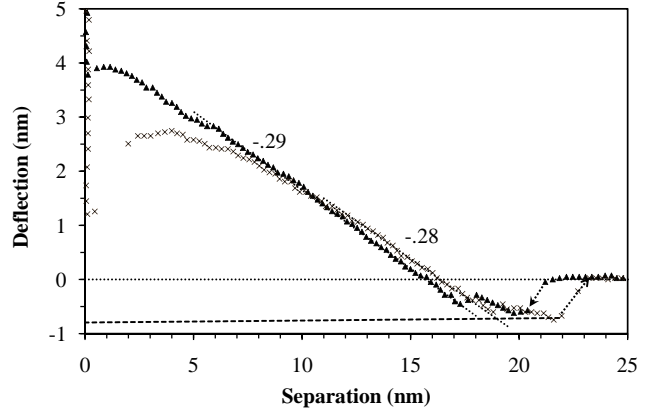} } }
\caption{\label{Fig:AFM-2F10}
Measured cantilever deflection versus separation
for a nanobubble
(curves and lines as in preceding figure;
different cantilever, solution, and nanobubble).
The cantilever has
spring constant $k_\mathrm{t}=0.35\,$N/m,
and conical half-angle $\alpha = 10^\circ\!$.
The undeformed nanobubble has measured
height $z_\mathrm{c}=19.5\,$nm
and substrate contact radius $r_3 =205\,$nm,
corresponding to a curvature radius $R_\mathrm{c} = 1087\,$nm
and a contact angle of $169^\circ$.
The dashed curve is equilibrium (slip) linear theory with
$r_0=20\,$nm, $\gamma = 0.037\,$N/m,
$s=1.68$, 
and $\Delta \gamma=+3.8\,$mN/m (equivalently, $\theta_\mathrm{t}=87^\circ$).
}
\end{figure}

Figure \ref{Fig:AFM-2F10}  shows yet another nanobubble measurement.
The AFM fluid cell was flushed with ethanol and then water
which means that there was likely exothermic heating.
The undeformed nanobubble has apex height $z_\mathrm{c}=19.5\,$nm
with the first indication of a force occurring at a separation of 21.9$\,$nm.
The nanobubble was re-imaged after 47 force measurements
and the apex height was $z_\mathrm{c}=21.5\,$nm,
and the substrate contact radius was $r_3 =215\,$nm,
which correspond to a curvature radius $R_\mathrm{c} = 1081\,$nm.
Again this is unambiguous evidence that the nanobubble
is thermodynamically stable.

The two linear regions measured in the figure,
one each on extension and retraction,
may be attributed to stick.
Assuming  a contact radius of $r_1=20\,$nm,
the slope $-0.28$ corresponds to $\gamma(s) = 0.037\,$N/m.
Alternatively for this slope,
$r_1=10\,$nm corresponds to $\gamma(s) = 0.048\,$N/m.
and  $r_1=50\,$nm $\gamma(s) = 0.022\,$N/m.
The exact and linear bubble spring constants agree to better than 0.05\%
in this regime.

The value $\gamma(s) = 0.037\,$N/m,
and the nanobubble radius of curvature $R_\mathrm{c} = 1070\,$nm
correspond to a supersaturation value of $s=1.7$.
Using this in the linear model for the supersaturated surface tension
gives a spinodal supersaturation ratio of $s^\ddag =2.4$.
Alternatively,
$\gamma(s) = 0.048\,$N/m
corresponds to  $s=1.9$ and  $s^\ddag =3.6$.

One can assume that the cantilever initially jumps in to the equilibrium slip
position,
which is supported by the flat nature of the deflection curve
and the coincidence of approach and retraction.
Assuming again a flattened conical tip with  $r_0=20\,$nm,
these data are well-fitted using $\gamma(s) = 0.037\,$N/m, $s=1.7$,
and $\Delta \gamma=+3.8\,$mN/m,
which would correspond to a macroscopic contact angle
of water on silicon nitride of $\theta_\mathrm{t}=87^\circ$.
The validity of the fit is supported by the coincidence
of the jump-out separation and the end of the stability
of the extended nanobubble.
It can be mentioned that a virtually identical equilibrium slip curve
can be obtained for  $r_0=50\,$nm,
with $\gamma(s) = 0.022\,$N/m, $s=1.41$,
and with $\Delta \gamma=+2.3\,$mN/m,
which would correspond to a macroscopic contact angle
of water on silicon nitride of $\theta_\mathrm{t}=88.2^\circ$.
The equilibrium data can also be fitted by
 $r_0=10\,$nm,
with $\gamma(s) = 0.048\,$N/m, $s=1.9$,
and with $\Delta \gamma=+2.5\,$mN/m, ($\theta_\mathrm{t}=88^\circ$).

\begin{figure}[t!]
\centerline{\resizebox{8.5cm}{!}{ \includegraphics*{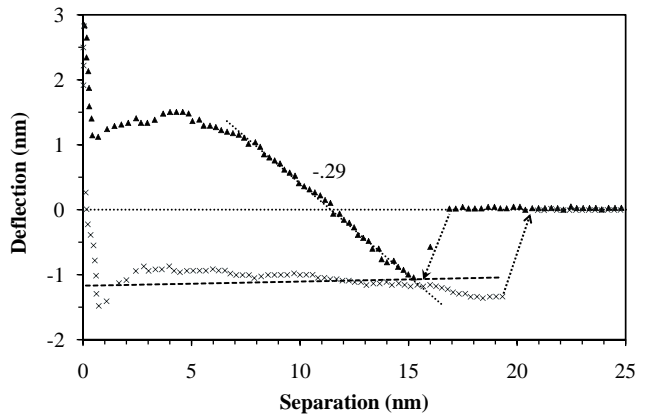} } }
\caption{\label{Fig:AFM-22F18}
Measured cantilever deflection versus separation
for a nanobubble
(same cantilever and solution  as in the preceding figure;
different nanobubble).
The undeformed nanobubble has measured
height $z_\mathrm{c}=15.9\,$nm
and substrate contact radius $r_3 =174\,$nm,
corresponding to a curvature radius $R_\mathrm{c} = 970\,$nm
and a contact angle of $170^\circ$.
The dashed curve is equilibrium (slip) linear theory with
$r_0=20\,$nm, $\gamma(s) = 0.035\,$N/m,
and $s=1.72$, 
and $\Delta \gamma=+2.5\,$mN/m (equivalently, $\theta_\mathrm{t}=88^\circ$).
}
\end{figure}

Figure \ref{Fig:AFM-22F18} shows another nanobubble
in the same solution as Fig.~\ref{Fig:AFM-2F10}.
Despite the differences in height and contact radii,
the two nanobubble have about the same radius of curvature
($R_\mathrm{c} = 1087\,$nm there and $970\,$nm here).
This is consistent with the level of supersaturation of the solution
being unchanged and with the nanobubbles being in diffusive equilibrium.
Likewise, one would expect the surface tension to be unchanged,
and this is confirmed by the fact that the slope of the linear pinned regions
are about the same ($-0.28$ there and $-.29$ here).

Using a contact radius of $r_0 = 20\,$nm,
the slope of the stick region in extension in Fig.~\ref{Fig:AFM-22F18} of
$-0.29$
corresponds to a surface tension of $\gamma = 0.035\,$N/m,
a supersaturation ratio of $s=1.72$,
and a spinodal supersaturation ratio of $s^\ddag =2.4$.
The minimum of the extension curve touches a calculated equilibrium slip curve
for $\Delta \gamma=+2.5\,$mN/m
(equivalently, $\theta_\mathrm{t}=88^\circ$).

Using instead a contact radius of $r_0 = 50\,$nm,
the slope of the stick region in extension in Fig.~\ref{Fig:AFM-22F18}
of $-0.29$ corresponds to
$\gamma(s) = 0.020\,$N/m, $s=1.42$, $s^\ddag=1.58$.
Fitting the minimum of  the extension curve
gives $\Delta \gamma=+1.3\,$mN/m (equivalently, $\theta_\mathrm{t}=89^\circ$).

Using instead a contact radius of $r_0 = 10\,$nm,
the slope of $-0.29$ corresponds to
$\gamma(s) = 0.046\,$N/m, $s=1.95$, $s^\ddag=3.61$.
Fitting the minimum of  the extension curve
gives $\Delta \gamma=+3.8\,$mN/m (equivalently, $\theta_\mathrm{t}=87^\circ$).

It should be mentioned that the slope of the stick region was checked
against that given by the exact theory and the agreement was better than 1\%.
The origin of the non-linearity and peak in the putative pinned region
in  Fig.~\ref{Fig:AFM-22F18} (and in Fig.~\ref{Fig:AFM-2F10}) is unclear,
although one could speculate that the contact line
might be moving with a finite velocity in these regions.

The slight negative slope in the putative equilibrium curve
here in Fig.~\ref{Fig:AFM-22F18}
is difficult to reproduce in the theoretical calculations.


%
\section{Conclusion}
\setcounter{equation}{0} 
%

\begin{table}[t]
\caption{\label{Tab:Nanob10}
Measured and Deduced Properties of Nanobubbles and Solutions
($r_0=10\,$nm).}
\begin{center}
\begin{tabular}{c c c c c c c c c}
\hline\noalign{\smallskip}
Figure &  $k_\mathrm{t}$ &
$r_3$ & $z_\mathrm{c}$ & $R_\mathrm{c}$ &
$\gamma(s)$ & $s$ & $s^\ddag$  & $\theta_\mathrm{t}$ \\
    &     (N/m)   & (nm)  &  (nm) & (nm) & (N/m) & & &  (deg.)\\
\hline \\
\ref{Fig:AFM-F3} & 0.35 & 108 & 33 & 192 & 0.040 & 5.2 & 10.4 & 98 \\
\ref{Fig:AFM-F7} & 0.24 & 82.5 & 9.6 & 359 & 0.041 & 3.3 & 6.3 & 89 \\
\ref{Fig:AFM-2F10}$^*$ & 0.35 & 205 & 19.5 & 1087 & 0.047 & 1.9 & 3.6 & 87 \\
\ref{Fig:AFM-22F18}$^*$ & 0.35 & 174 & 15.9 & 970 & 0.046 & 2.0 & 3.6 & 88 \\
\hline
\end{tabular} \\
\flushleft
$^*$ Same solution, different nanobubble.
\end{center}
\end{table}

\comment{
\begin{table}[t]
\caption{\label{Tab:Nanob20}
Measured and Deduced Properties of Nanobubbles and Solutions
($r_0=20\,$nm).}
\begin{center}
\begin{tabular}{c c c c c c c c c}
\hline\noalign{\smallskip}
Figure &  $k_\mathrm{t}$ &
$r_3$ & $z_\mathrm{c}$ & $R_\mathrm{c}$ &
$\gamma(s)$ & $s$ & $s^\ddag$  & $\theta_\mathrm{t}$ \\
    &     (N/m)   & (nm)  &  (nm) & (nm) & (N/m) & & &  (deg.)\\
\hline \\
\ref{Fig:AFM-F3} & 0.35 & 108 & 33 & 192 & 0.030 & 4.1 & 6.2 & 96 \\
\ref{Fig:AFM-F7} & 0.24 & 82.5 & 9.6 & 359 & 0.028 & 2.5 & 3.5 & 88 \\
\ref{Fig:AFM-2F10}$^*$ & 0.35 & 205 & 19.5 & 1087 & 0.037 & 1.68 & 2.38 & 87 \\
\ref{Fig:AFM-22F18}$^*$ & 0.35 & 174 & 15.9 & 970 & 0.035 & 1.72 & 2.40 & 88 \\
\hline
\end{tabular} \\
\flushleft
$^*$ Same solution, different nanobubble.
\end{center}
\end{table}
} 

The present paper gives analytic expressions for the nanobubble
spring constant that allow its surface tension to be obtained
from the slope of the pinned regions in a force-separation
AFM measurement.
Expressions are also given that allow the difference in tip surface energies
(tip contact angle) to be obtained from an equilibrium part
of the force curve.

The present fits to the experimental data do not give enough information
to pin down the value of the blunt tip radius $r_0$.
However, sensible results are generated
by assuming a tip and contact radius $r_0 = r_1 = 10\,$nm
(Table~\ref{Tab:Nanob10}).
This is at the lower end of the range
of a typical AFM tapping mode cantilever tip,
20--60$\,$nm.
A high value of the radius,  $r_0 = r_1 = 50\,$nm
gave quite low values of the surface tension,
supersaturation ratio, and spinodal supersaturation ratio.

The most reliable data appears to be that of Fig.~\ref{Fig:AFM-F3}.
In this case the two pinned regions both appear to begin
from the equilibrium curve,
which means that the change in contact radius
can be found from the change in separation of the starts.
Hence one has three knowns (the two slopes and the change in contact radius)
and three unknowns (the surface tension and the two contact radii).
Solving this system yields the contact radii $r_1=9.4\,$nm and $r_1'=11.4\,$nm,
and also
$\gamma(s) = 0.041\,$N/m, $s=5.2$ and $s^\ddag=10.8$.
Unfortunately the other figures do not have pinned regions
starting from the equilibrium curve and so the change in contact radius
cannot be readily deduced.

The estimates of the surface energy difference
and macroscopic tip contact angle
are based on the equilibrium curves
and are not so reliable.
Little more can be said than that the contact angle is close to $90^\circ$.

Likewise the estimate of the value of the spinodal supersaturation
ratio has limited reliability because of the simplicity
of the linear supersaturated surface tension model.
It is nonetheless consoling that it comes out to be
of the same order as has been found in computer simulations
of a Lennard-Jones fluid.\cite{Moody03,Moody04,He05}

One of the purposes of this study was to establish experimentally
that the surface tension of nanobubbles
was less than that of saturated water.
The surface tension is obtained from the slope of the pinned regions
and does not rely upon the surface energy difference
nor  the spinodal supersaturation ratio.
The greatest uncertainty concerns the tip radius,
and to this end it is better to use a low value,
since this overestimates the surface tension.
(Doing the calculations at constant chemical potential
rather than at constant number further overestimates the surface tension.)
Hence the data in Table~\ref{Tab:Nanob10} for $r_0=10\,$nm,
which give $ 0.04 \alt \gamma(s) \alt 0.05 \,$N/m,
are most likely an upper bound on the nanobubble surface tension.
The nanobubble surface tension
is substantially reduced from that of the saturated air water interface,
$\gamma^\dag = 0.072\,$N/m.

The solution supersaturation ratios are deduced to be in the range
$1.9 \alt s \alt 5.2$ using  $r_0=10\,$nm.
These values  appear realistic
given the fact that
a 15$^\circ$C change in temperature is enough to change
the solubility of CO$_2$ by a factor of two.


It is clear that
in order to reliably obtain the dependence
of the surface tension on the supersaturation ratio
one needs to know the contact radius to within a few nanometers.
In contrast,
in order to prove that the solution is supersaturated
and that the surface tension is less than the saturated value
one does not need to know the tip radius precisely.

On this basis one can conclude that
the present analysis of these experimental measurements
on nanobubbles explicitly confirm
what is required by thermodynamics:
for nanobubble equilibrium the solution must be supersaturated,
and a supersaturated solution has a lower surface tension
than a saturated solution.\cite{Moody02,Moody03}

In future experimental  studies,
electron micrography or AFM inverse imaging
could be used to get an independent estimate of the cross-sectional
radius of the tip.
Also, attempts could be made
to explicitly control or to measure the supersaturation of the solution.

\emph{Acknowledgement.}
The AFM data used here was supplied by Liwen Zhu.
I thank her and Chiara Neto for interesting discussions
during an earlier stage of the project.





%
%


\end{document}